\documentclass{article}
\usepackage[utf8]{inputenc}

\usepackage{geometry}
\usepackage[square,sort,comma,numbers]{natbib}
\usepackage{amsmath}
\usepackage{latexsym}
\usepackage{amsfonts}
\usepackage{mathrsfs}
\usepackage{upgreek}
\usepackage{graphics}
\usepackage{graphicx}
\usepackage{epstopdf}
\usepackage{epsfig}
\usepackage{amsbsy}
\usepackage{amsfonts}
\usepackage{amsmath}
\usepackage{amssymb}
\usepackage{bm}
\usepackage{color}
\usepackage{setspace}
\usepackage{authblk}

\title{A continuous fracture front tracking algorithm with multi layer tip elements (MuLTipEl) for a plane strain hydraulic fracture}

\author{E.V. Dontsov$^1$}
\date{%
    $^1$egor@resfrac.com, ResFrac Corporation, 555 Bryant Street, \#185
    Palo Alto, CA 94301, USA
}

\begin{document}

\maketitle
\begin{abstract}
    \noindent The problem of a plane strain hydraulic fracture propagating in a layered formation is considered. Fracture toughness, \emph{in-situ} stress, and leak-off coefficient are assumed to vary by layer, while the elastic properties are kept constant throughout the domain for simplicity. The purpose of this study is to develop a numerical algorithm based on a fixed mesh approach, which is capable to solve the above problem accurately using elements which can even be larger than the layer size. In order to do this, the concept of fictitious tip stress is first introduced for determining the fracture front location. In this technique, an additional stress is applied to the tip element with the purpose to suppress opening and to mimic width corresponding to the actual fracture front location. A theoretical basis for this concept has been established and it is further calibrated for piece-wise constant elements. Once the ability to track the crack front location is developed, the effect of layers is included by vary properties as a function of front location. Several numerical examples benchmarking the numerical solution, as well as highlighting capabilities of the algorithm to tackle multiple thin layers accurately are presented.
\end{abstract}

\section{Introduction}

Hydraulic fractures are often generated underground to stimulate rock formations for the purpose of hydrocarbon extraction~\cite{Econo2000,King2012}. In the context of low-permeability shale formations, multiple wells are typically drilled close to each other, which are then fractured in stages obeying a certain sequence. Each stage consists of multiple entry points or perforation clusters, which leads to simultaneous generation of multiple hydraulic fractures. In addition to that, rock formations are highly laminated and the layers can be distinguished virtually on any length scale. The latter raises a fundamental question of how can we upscale the properties and to what resolution, and how can we run a numerical simulator on a high resolution data? The first part of this question has been addressed in~\cite{Dont2020c}, while this study focuses on the second one, namely, how to properly use high resolution rock properties in a hydraulic fracturing model. To moderate complexity of the analysis, the canonical case of a single plane strain hydraulic fracture is considered.

In order to incorporate high resolution data into the algorithm using a coarse mesh, it is first necessary to develop an ability to track the fracture front. The vast majority of numerical simulators for modeling hydraulic fracturing assume opening of one element at a time, see e.g.~\cite{Damj2013,Xu2013,Sher2015,Shiozawa2016,Baykin2018,Dont2018b,Reza2020}. Such approaches are not able to explicitly account for the effect of thin layers that are smaller than the element size because only a single set of properties is assigned to every element. At the same time, it is possible to use homogenization~\cite{Dont2020c} to implicitly account for the effect of thin layers. But the accuracy of fracture front location is still determined approximately up to an element size, which can introduce significant errors if a relatively large elements are used. 

Some methods use a moving mesh approach, which allows to track gradual evolution of the fracture front relative to layers and effectively have several layers per element~\cite{Dont2017j}. In the latter study, solution for tip element that includes multiple layers is computed numerically for every time step and is therefore the overall approach is not computationally efficient. Pseudo-3D models~\cite{Sett1986,Kress2013,Cohen2015,Dont2015c} can also be conditionally included into this category since the fracture height is tracked ``smoothly'' for these and the effect of layers is included. The primary disadvantages of the aforementioned moving mesh approaches are the necessity to constantly update element locations relative to layers and also the extension to a fully planar or multi-planar model presents significant challenges.

Another possibility is to track fracture front, now with a fixed mesh, is to use an Implicit Level Set Algorithm (ILSA), pioneered in~\cite{Peir2008}, and then further extended in~\cite{Peir2015,Dont2017a,Zia2019}. This algorithm is based on explicitly incorporating the near tip asymptotic solution into the numerical scheme as a propagation condition. In particular, for the case of a plane strain fracture, the fracture front location and width of the tip element are determined by the width of the penultimate element or so-called survey element. However, such a ``non-locality'' of the boundary condition may present serious challenges when dealing with thin layers and it also requires a good approximation for the near tip solution that accounts for layers. 

While it is probably possible to utilize ILSA approach mentioned above to solve the problem under consideration, to provide a potentially simpler alternative, another methodology that uses ``local'' boundary condition is developed. In this method, the partially filled tip element is completely independent from its fully open neighbour and its width is used to determine the fracture front location. Arguably, one the biggest challenges with such an approach is to accurately capture the toughness dominated regime, or uniformly pressurized fracture. Once a new element is introduced, then, under uniform pressure condition, the solution immediately ``jumps'' to the one corresponding to length increased by one element. To prevent this from happening, an additional fictitious stress is added to the tip element. Indeed, if the additional stress is sufficiently large, then the newly introduced element remains closed and the apparent fracture length remains the same. At the same time, as the fictitious stress decreases to zero, the solution transitions to the one with the increased length. Thus, this gives an opportunity to smoothly evolve the apparent fracture length, and hence to track the fracture front by gradually changing magnitude of the additional stress at the tip element. This concept lays the foundation for the algorithm presented here.

The organization is the following. Section~\ref{SecProb} outlines the governing equations for the plane strain hydraulic problem under consideration. Then, Section~\ref{SecFicStress} presents theoretical basis for the algorithm as well as calibrates it for the case of piece-wise constant elements. This section also describes the methodology to include the effects of fluid viscosity, leak-off, and layers. After this, Section~\ref{SecNum} presents a series of numerical results that evaluate accuracy of the developed approach against the existing reference solutions and also illustrates ability of the algorithm to provide mesh independent results for thinly layered rock properties. Finally, Section~\ref{SecSum} summarizes the main findings.

\section{Problem formulation}\label{SecProb}

\begin{figure}
\centering \includegraphics[width=0.5\linewidth]{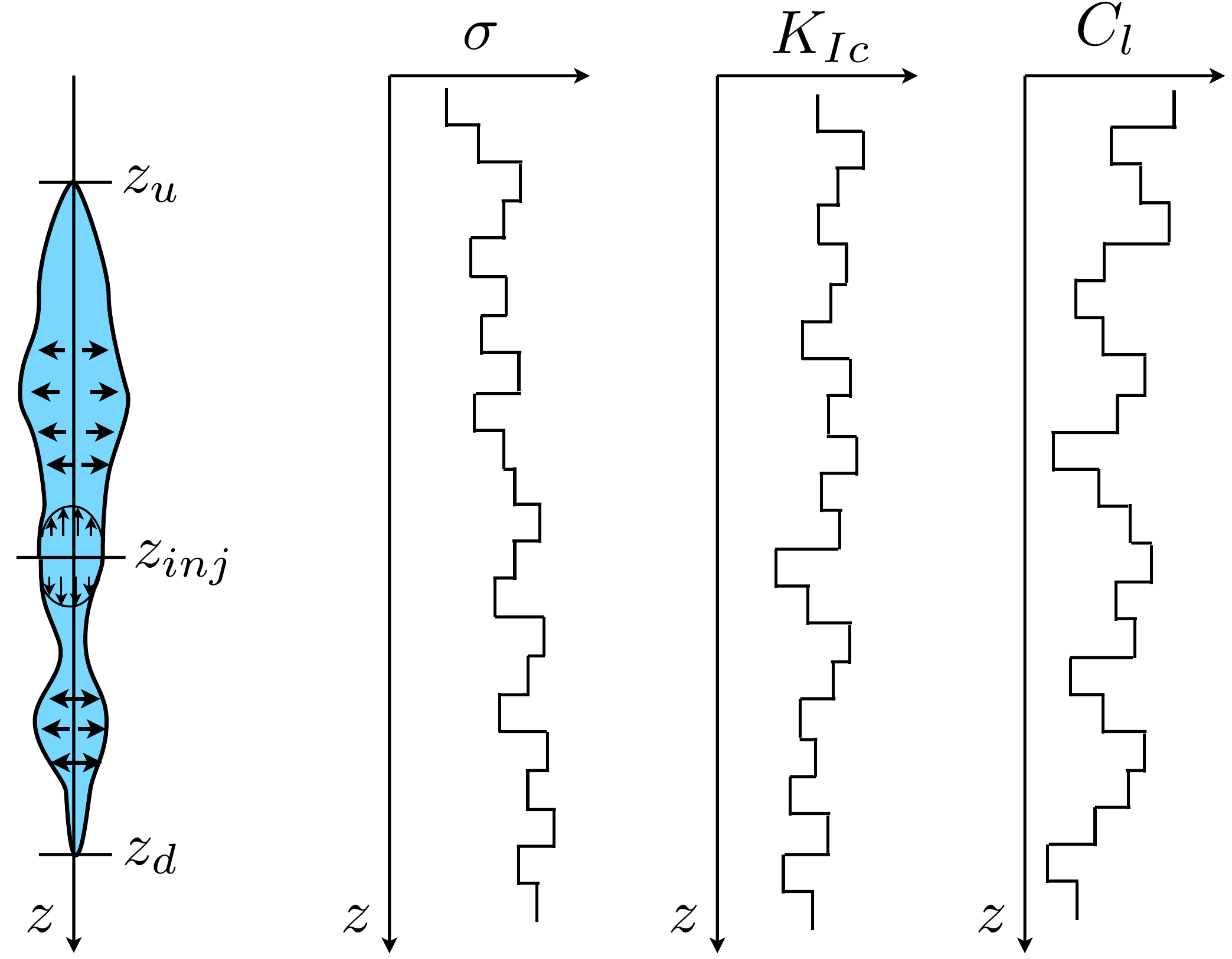}
\caption{Schematics of a plane strain hydraulic fracture propagating in a layered formation, in which stress, toughness, and leak-off vary with depth.}
\label{fig1}
\end{figure}

Consider a vertical plane strain hydraulic fracture that propagates along the depth or $z$ direction in a layered formation, as depicted in Fig.~\ref{fig1}. It is assumed that the fracture is driven by a Newtonian fluid with viscosity $\mu$ that is injected with a constant the rate $Q_0$. Elastic properties are assumed to be homogeneous and characterized by Young's modulus $E$ and Poisson's ratio $\nu$. In contrast, fracture toughness $K_{Ic}$, stress $\sigma$, and Carter's leak-off coefficient $C_l$ are assumed to vary from layer to layer.
To reduce lengthy expressions, the following scaled parameters are introduced:
\begin{equation}
    E' = \dfrac{E}{1-\nu^2},\qquad K' = \sqrt{\dfrac{32}{\pi}}K_{Ic}, \qquad \mu' = 12\mu, \qquad C' = 2C_l.
\end{equation}


Volume balance for an incompressible Newtonian fluid inside the one-dimensional crack can be written as
\begin{equation}\label{volumebalance}
\dfrac{\partial w}{\partial t}+\dfrac{\partial q}{\partial z}+\dfrac{C'(z)}{\sqrt{t-t_0(z)}}=\dfrac{Q_0}{H}\delta(z\!-\!z_{inj}),\qquad q=-\dfrac{w^3}{\mu'}\dfrac{\partial p}{\partial z},
\end{equation}
where $w(z,t)$ denotes the fracture width, $q(z,t)$ is the flux along the crack (in the $z$ direction), $p(z,t)$ is the fluid pressure, $C'(z)$ is the scaled Carter's leak-off coefficient that varies with depth, and $Q_0$ is the fluid injection rate per unit length, which is located at the point $z_{inj}$. 

The elasticity equation for the plane strain fracture in an elastically homogeneous material is given by the following boundary integral formulation (see e.g.~\cite{Crouch1983, Hills1996})
\begin{equation}\label{elas}
p(z,t)=\sigma(z)-\dfrac{E'}{4\pi }\int_{z_u}^{z_d} \dfrac{1}{s-z}\dfrac{\partial w(s,t)}{\partial s}\,ds,
\end{equation}
where $z_u$ and $z_d$ are locations of the fracture front in the upwards and downwards directions and $\sigma(z)$ denotes the \emph{in-situ} stress that varies with depth. 

Fracture propagation condition is governed by the linear elastic fracture mechanics~\cite{Rice1968}
\begin{equation}\label{propcond}
w\rightarrow \dfrac{K'(z)}{E'}(z\!-\!z_u)^{1/2},\qquad z\rightarrow z_u,\qquad
w\rightarrow \dfrac{K'(z)}{E'}(z_d\!-\!z)^{1/2},\qquad z\rightarrow z_d,
\end{equation}
which states that the mode I stress intensity factor is equal to the fracture toughness for a propagating fracture. In addition to the propagation condition~(\ref{propcond}), zero flux conditions at the fracture tips are enforced, i.e. $q(z_u,t)\!=\!0$ and $q(z_d,t)\!=\!0$.

\section{Fictitious stress to model fracture evolution}\label{SecFicStress}

One of the challenges in modeling hydraulic fracture propagation is to have an ability to simulate gradual fracture front evolution. The simplest way is to open one element at a time. This, however, introduces mesh dependence and requires the element size to be smaller than the layer size to capture the effect of layering accurately. Therefore, there is a need to model gradual fracture front evolution, so that the fracture front can be located inside the element. 

\subsection{Theoretical concept}

\begin{figure}
\centering \includegraphics[width=0.7\linewidth]{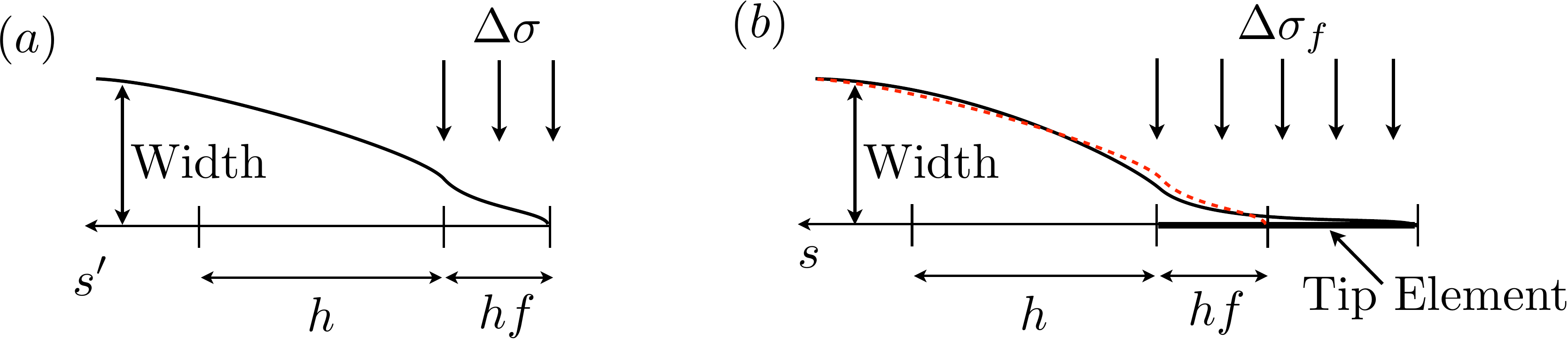}
\caption{Illustration of the fictitious stress concept. Panel $(a)$ schematically shows the exact solution with fracture front located within an element. Panel $(b)$ shows the fictitious solution, in which the fracture front is located at the edge of an element. The dashed red line shows the exact solution for comparison.}
\label{fig2}
\end{figure}

With the reference to Fig.~\ref{fig2}$(a)$, consider the situation in which the fracture propagates in relation to the existing mesh with size $h$. The actual fracture tip is located somewhere within the tip element and the fill ratio, $f$, is defined as the fraction of the element that is filled with fluid. In general, different elements may have different value of stress. Therefore, it is assumed that the tip element has an additional \emph{in-situ} stress $\Delta\sigma$ relative to its neighbour. To simulate smooth fracture front evolution in the numerical scheme, the concept of fictitious stress is introduced. As shown in Fig.~\ref{fig2}$(b)$, the tip element cannot be partially open numerically, so it has to be open fully. This in turn changes the fracture front location and may affect solution away from the tip. To compensate for this and to ensure that the far-field behavior is preserved, fictitious stress $\Delta\sigma_f$ is applied to the tip element, which is different from $\Delta\sigma$. Dashed red line in Fig.~\ref{fig2}$(b)$ schematically shows the exact solution, while the solid black line illustrates the solution with the fictitious stress. The behavior near the tip does not precisely match between the two solutions, while the far-field behavior is similar.

To better understand the relation between the fictitious stress $\Delta \sigma_f$ and the fill ratio $f$, the example of toughness dominated semi-infinite fracture is considered first. The fracture width for the problem with a stress barrier in the toughness regime is given by the following expression~\cite{Dont2017j}:
\begin{equation}\label{widthexact}
    w_{e} = \dfrac{K'}{E'}s'^{1/2}+\dfrac{4 \Delta \sigma}{\pi E'} F(s',hf),\qquad s' = s-(1\!-\!f)h\geqslant 0,
\end{equation}
where $s'$ is the distance from the tip and the function $F$ is given by
\[
F(s,x) = (s-x)\log\biggl|\dfrac{s^{1/2}+x^{1/2}}{s^{1/2}-x^{1/2}} \biggr|+2s^{1/2}x^{1/2}.
\]
Similar solution can be used for the problem with fictitious stress and different fracture front location:
\begin{equation}\label{widthfict}
    w_{f} = \dfrac{K'_t}{E'}s^{1/2}+\dfrac{4 \Delta \sigma_f}{\pi E'} F(s,h),\qquad 
\end{equation}
where $K'_f$ is the apparent toughness for the ``fictitious'' fracture at the tip. The far field behavior of the two solutions can be matched by requiring that
\begin{equation}\label{Kmatch}
 K'+\dfrac{16}{\pi}\Delta \sigma f^{1/2} h^{1/2} = K'_f+\dfrac{16}{\pi}\Delta \sigma_f h^{1/2},
\end{equation}
where it is used that $\lim_{s\rightarrow\infty}F(s,x)=4s^{1/2}x^{1/2}$. Further, the volume of the tip element can be matched as
\begin{equation}\label{Volmatch}
\dfrac{2K'f^{3/2} h^{1/2}}{3E'}+\dfrac{4 \Delta \sigma}{\pi E'h} \int_0^{hf} F(s',hf)\,ds' = \dfrac{2K_f' h^{1/2}}{3E'}+\dfrac{4 \Delta \sigma_f}{\pi E'h} \int_0^h F(s,h)\,ds.
\end{equation}
Given that the integrals in the above expression can be computed as $\int_0^x F(s,x)\,ds=2x^2/3$, equation~(\ref{Volmatch}) can be further simplified using~(\ref{Kmatch}) to
\begin{equation}\label{Sigmaf}
 \Delta\sigma_f = \dfrac{K'} {h^{1/2}}\Sigma_K +\Delta\sigma \Sigma_S,\qquad \Sigma_K = \dfrac{\pi}{12}(1\!-\!f^{3/2}),\qquad \Sigma_S = \dfrac{f^{1/2}}{3}(4\!-\!f^{3/2}).
\end{equation}
This result indicates that the fictitious stress has two contributions, the first one is due to toughness, and the second one caused by the stress barrier. The toughness contribution decays gradually as the fill ratio increases and it vanishes once $f=1$. In contrast, contribution of stress is zero for $f=0$ and it reaches one for $f=1$. This result makes sense since in the limiting case of $f=0$, the fracture is not exposed to the stress barrier (hence it does not affect the solution), but there should be an additional stress to prevent the newly created element from opening. At the same time, for $f=1$ there should be no contribution due to toughness since the fracture front is located at the edge of an element and the stress contribution is essentially equal to $\Delta\sigma$.

The above considerations are based on the analytical solution and may not necessarily apply exactly in the numerical scheme, in which piece-wise constant elements with size $h$ (comparable to the problem dimensions) are used. Therefore, the exercise needs to be repeated in the context of the numerical solution with the goal to compute the expressions for $\Sigma_K$ and $\Sigma_S$, similar to~(\ref{Sigmaf}). 

\subsection{Correction for numerical discretization}

This section aims to repeat the concept outlined in the previous section for a finite length fracture for the purpose of accounting for the discretization errors. In order to do this, analytical solution for a uniformly pressurized finite fracture with symmetric stress barriers is compared to the numerical solution.

Consider a plane strain uniformly pressurized fracture in a homogeneous formation. Let the scaled fracture toughness be $K'$, element size be $h$, and length of a single fracture wing be $L=(N\!+\!f)h$, where $f$ is the fill ratio and $N$ is the number of fully open elements for a single fracture wing. Also, let the additional stress $\Delta\sigma$ be applied to the last element. Analytical solution for the fracture width for such a problem is given by~(see e.g.~\cite{Adachi2010,Dont2015c})
\begin{equation}\label{widthexact2}
    w = \dfrac{K'}{E'}\dfrac{\chi}{\sqrt{2 L}}+\dfrac{4\Delta\sigma}{\pi E'}\biggl\{-z\log\biggl|\dfrac{L_0\chi+z\psi}{L_0\chi-z\psi}\biggr|+L_0\log\biggl|\dfrac{\chi+\psi}{\chi-\psi}\biggr|\biggr\},
\end{equation}
where $L_0=Nh$, $\chi = \sqrt{L^2-z^2}$, $\psi=\sqrt{L^2-L_0^2}$, and $0\leqslant z\leqslant L$. The fracture volume is
\begin{equation}
    V = \dfrac{\pi}{\sqrt{8}}\dfrac{K'}{E'} L^{3/2}+\dfrac{4L_0}{E'\Delta\sigma}\sqrt{L^2-L_0^2}.
\end{equation}
The width of the tip element can be computed according to~(\ref{Volmatch}) as
\begin{equation}\label{wtip}
  w_{tip} = \dfrac{2K'}{3E'}f^{3/2} h^{1/2}+\dfrac{8\Delta\sigma h f^2}{3\pi E'}.
\end{equation}

To construct the numerical solution, piece-wise constant width approximation is employed. In this case, the elasticity equation~(\ref{elas}) for a uniformly pressurized fracture with $N+1$ open elements and fictitious stress at the tip can be written in the discretized form as
\begin{equation}\label{wNum}
    \boldsymbol{w} = \boldsymbol{C}^{-1} \boldsymbol{v_1} p_0 +  \boldsymbol{ C}^{-1} \boldsymbol{v_2} \Delta\sigma_f,
\end{equation}
where $\boldsymbol{w}$ is an array of fracture widths, $\boldsymbol{C}$ is the elasticity matrix (see Appendix~\ref{app2} for the exact expression), $\boldsymbol{v_1}$ is the vectors of ones, $p_0$ is the fracture pressure, $\boldsymbol{v_2}$ is the vectors of zeros, in which the last term corresponding to the tip element is equal to one, and $\Delta \sigma_f$ is magnitude of the fictitious stress. Note that only half of the domain is considered due to symmetry. To ``match'' the analytical and numerical solutions, the total fracture volume and the tip element volume need to be made equal, so that
\begin{equation}\label{voltipmatch}
    \sum \boldsymbol{w} h = V,\qquad w_{N+1} = w_{tip},
\end{equation}
where $w_{N+1}$ is the width of the tip element computed numerically. Solution of the equations~(\ref{wNum}) and~(\ref{voltipmatch}) can be written as
\begin{equation}\label{sigmasolnum}
    \Delta\sigma_f = \dfrac{K'}{h^{1/2}}\Sigma_K(f,N)+\Delta\sigma \Sigma_S(f,N),
\end{equation}
see Appendix~\ref{app1} for details. Functions $\Sigma_K(f,N)$ and $\Sigma_S(f,N)$ are computed numerically.

\begin{figure}
\centering \includegraphics[width=0.7\linewidth]{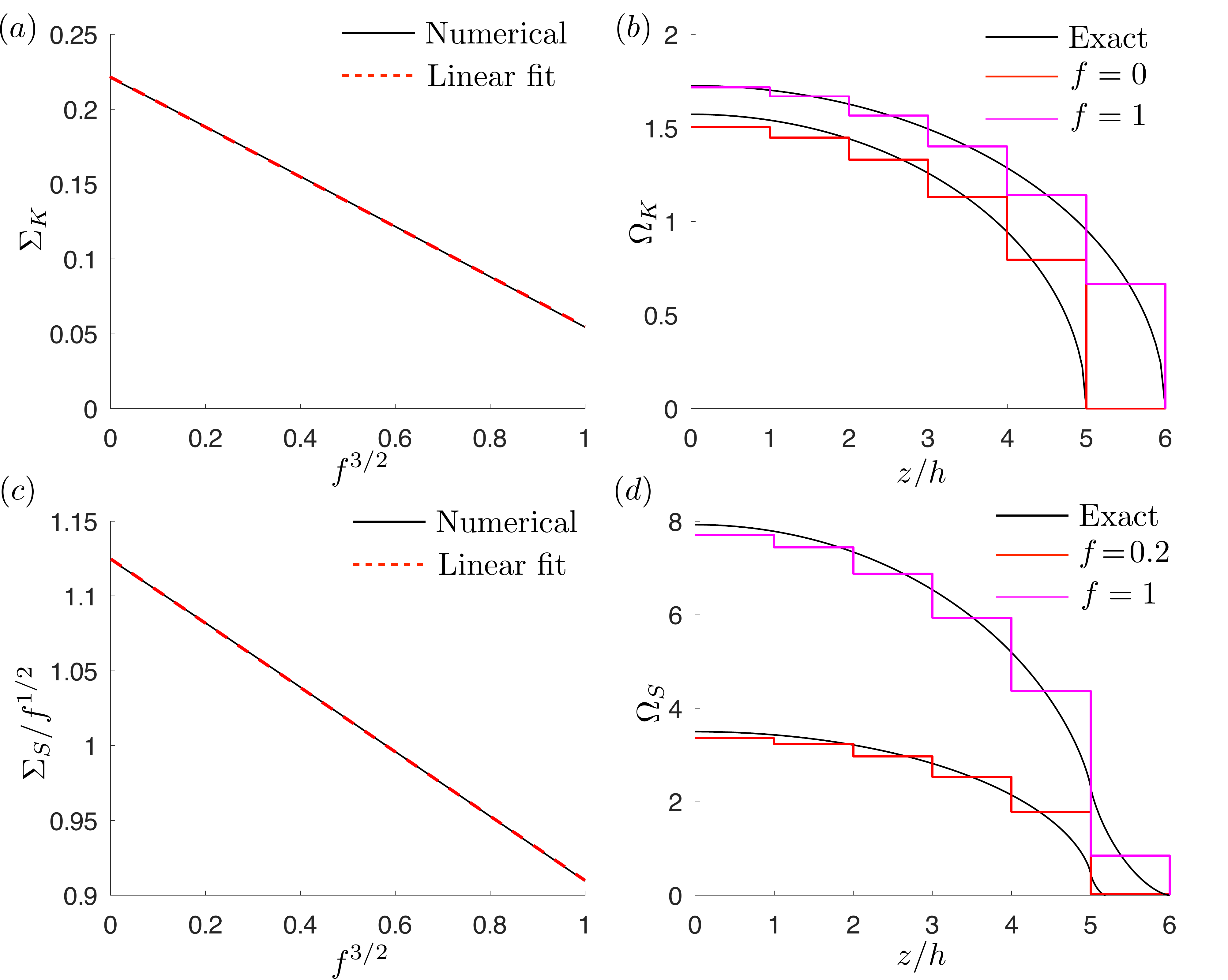}
\caption{Panel $(a)$: Variation of the dimensionless fictitious stress due to toughness versus fill ratio to the power of $3/2$ (black solid line) for $N=50$ fracture elements. The red dashed line shows the linear fit~(\ref{sigmaKfit}). Panel $(b)$: Comparison between the exact (black lines) and numerically computed fracture width for $f=0$ (red line) and $f=1$ (magenta line) for the case of $N=5$ fully open elements and $\Delta\sigma=0$. Panel $(c)$: Variation of the dimensionless fictitious stress due to stress barrier, normalized by $f^{1/2}$, versus fill ratio to the power of $3/2$ (black solid line) for $N=50$ fracture elements. The red dashed line shows the linear fit~(\ref{sigmaSfit}). Panel $(d)$: Comparison between the exact (black lines) and numerically computed fracture width for $f=0.2$ (red line) and $f=1$ (magenta line) for the case of $N=5$ fully open elements and $K'=0$.}
\label{fig3}
\end{figure}

Fig.~\ref{fig3}$(a)$ shows variation of the dimensionless fictitious stress $\Sigma_K$ versus $f^{3/2}$ for $N\!=\!50$ computed numerically (solid black line) as well as linear fit shown by the dashed red line. The linear trend is matched by using the equation
\begin{equation}\label{sigmaKfit}
    \Sigma_K = 0.221-0.167 f^{3/2}. 
\end{equation}
Note that the fitting parameters are the same to the third digit for $N=\{5,50,500\}$, so that there is practically no mesh dependence. However, some visible deviations start to occur for $N\!<\!5$. 
Comparison with~(\ref{Sigmaf}) demonstrates that the numerically computed fictitious stress has similar behavior, but quantitatively is different, and, in particular, it does not vanish for $f\!=\!1$.

Fig.~\ref{fig3}$(b)$ shows dimensionless width $\Omega_K = w E'/(K'h^{1/2})$ (no stress barrier, i.e. $\Delta\sigma\!=\!0$) for the two limiting cases of $f=0$ (solid red line) and $f=1$ (solid magenta line) for $N\!=\!5$, as well as the exact analytical solution for these cases~(\ref{widthexact}) (solid black lines). The case with $f=0$ demonstrates that even if the tip element is created, the additional fictitious stress is able to keep it closed. Also note a slight difference in the ``matching quality" to the exact solution for the cases with $f=0$ and $f=1$. The case with a fully filled tip matches much better. The difference comes from the fact that an additional non-zero stress is acting on the tip element for $f=1$. At the same time, there is stress at the tip element for $f=0$, but this element has zero width and therefore is practically inactive. The last relevant element is penultimate element. But it has no additional stress, and this is the source of the discrepancy with the solution for $f=1$ and also with the exact solution. The slight discrepancy of matching quality to the exact solution for $f=0$ and $f=1$ will manifest itself during fracture propagation. Once a new element is open, it will cause an immediate small width drop, even though the tip element will remain closed. One possible solution is to also apply stress to the penultimate element, but in this case the extension to planar fracture can become problematic.

Fig.~\ref{fig3}$(c)$ shows variation of the normalized dimensionless fictitious stress $\Sigma_S/f^{1/2}$ versus $f^{3/2}$ for $N\!=\!50$ (this case corresponds to zero toughness or $K'\!=\!0$). Solid black line shows the numerical solution, while the linear fit is shown by the dashed red line. The corresponding fitted expression for $\Sigma_S$ is
\begin{equation}\label{sigmaSfit}
    \Sigma_S = f^{1/2}(1.128-0.212 f^{3/2}). 
\end{equation}
In contrast to the previous case, the fitting parameters have some variation for $N=\{5,50,500\}$, on the order of 10\%. The equation above summarizes the case with the largest number of points.
Comparison between the theoretical prediction~(\ref{Sigmaf}) and~(\ref{sigmaSfit}) shows that the numerically computed fictitious stress has similar behavior, but has different numeric coefficients.

Fig.~\ref{fig3}$(d)$ compares the dimensionless width $\Omega_S = wE'/(\Delta\sigma h)$ for the exact solution~(\ref{widthexact}) and numerical solution with $N\!=\!5$. The exact solution is shown by the solid black lines, the numerical solution for $f=0.2$ is shown by the red line, while the numerical solution for $f=1$ is shown by the magenta line. Results demonstrate a good degree of agreement, including the volume of the tip element.

To summarize, this section demonstrates that in order to describe ``smooth'' propagation of a toughness dominated fracture in the context of piece-wise constant elements, the fill ratio needs to be updated based on the tip element width using~(\ref{wtip}). Then, once the fill ratio is computed, the fictitious stress needs to be computed using~(\ref{sigmasolnum}). In the latter equation, the numerically computed functions~(\ref{sigmaKfit}) and~(\ref{sigmaSfit}) should be used to ensure accuracy of the result for the case of constant displacement discontinuity elements.

\subsection{Correction for fluid viscosity and leak-off}

The above results are obtained for the case of toughness dominated fracture propagation, or assuming a uniformly pressurized fracture. Situation changes for the case of fluid driven fracture since the pressure distribution is no longer uniform and the fracture shape changes. However, for the case of homogeneous formation (i.e. no layers) the tip asymptotic solution is known~\cite{Dont2015d,Dont2018,Bessm2019} and can be written as
\begin{equation}\label{tipwidthgeneral}
w = \dfrac{K'\tilde w(\tilde s,\chi)}{E'} s^{1/2},
\end{equation}
where $s$ is the distance from the tip, while the function $\tilde w$ has the meaning of an apparent toughness and has an explicit expression for Newtonian, power-law, and Herschel-Bulkley fluids~\cite{Dont2015d,Dont2018,Bessm2019}. For the case of Newtonian fluid, it depends on
\begin{equation}
    \tilde s = \Bigl(\dfrac{s}{l}\Bigr)^{1/2},\qquad l = \Bigl(\dfrac{K'^{3}}{E'^{2}\mu'\dot s}\Bigr)^{2},\qquad \chi = \dfrac{2C'E'}{\dot s^{1/2} K'},
\end{equation}
which in turn depend on rock properties $E'$, $K'$, $C'$, fluid viscosity $\mu'$, distance from the tip $s$, and tip velocity $\dot s$. 

Using the result in~(\ref{tipwidthgeneral}), the expression for the average width of the tip element~(\ref{wtip}) and the associated fictitious stress~(\ref{sigmasolnum}) can be generalized to
\begin{equation}\label{bc_visc}
      w_{tip} = \dfrac{2K'\tilde w(f,f_0)}{3E'}f^{3/2} h^{1/2}+\dfrac{8\Delta\sigma h f^2}{3\pi E'},\qquad \Delta\sigma_f = \dfrac{K'\tilde w(f,f_0)}{h^{1/2}}\Sigma_K(f)+\Delta\sigma \Sigma_S(f),
\end{equation}
The function $\tilde w$ depends on material parameters and tip velocity, which can be written as $\dot s=h(f\!-\!f_0)/\Delta t$. Here $f_0$ is the fill ratio at the previous time step and $\Delta t$ is magnitude of the time step. Hence, to highlight the dependence on the previous fill ratio, it is written as $\tilde w(f,f_0)$. 

Even though the equation~(\ref{bc_visc}) represents an ``ad-hoc'' correction, it will be shown by the numerical examples that it is able to capture gradual fracture front evolution accurately even for viscosity dominated regime, for which the apparent toughness $\tilde w$ dominates the response. In the numerical algorithm, once the tip width is known, the first equation in~(\ref{bc_visc}) is solved for the new fill ratio $f$ and then the stress is computed using the second equation in~(\ref{bc_visc}).

\subsection{Incorporation of the effects of thin layers}

\begin{figure}
\centering \includegraphics[width=0.7\linewidth]{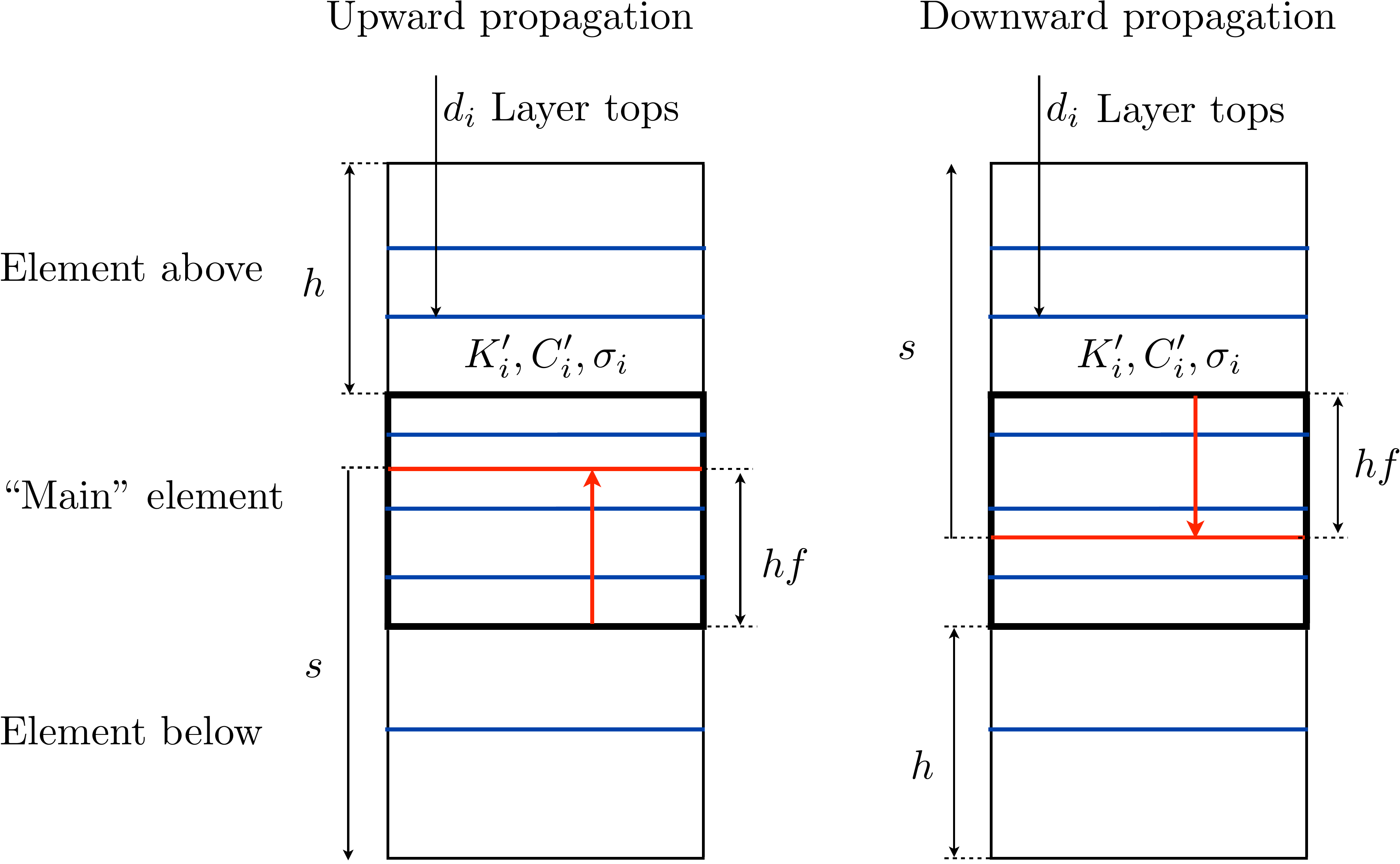}
\caption{Schematics of the upward and downward fracture propagation. The central ``main'' element is highlighted by the bold black perimeter and is surrounded by the elements above and below. Layers are shown by the blue lines. Fracture front is shown by the red line.}
\label{fig4}
\end{figure}

To include the effects of thin layers on fracture propagation, it is first necessary to gather this information from the layered rock properties. As indicated in Fig.~\ref{fig4}, each element needs to contain data for itself, as well as the element above and below. The ``main'' or central element is highlighted by the bold black line, while the layer boundaries are shown by the solid blue lines. Let $d_i$ denote depth of the $i$th layer top, such that the $i$th layer is located between $d_i$ and $d_{i+1}$ and is characterized by toughness $K'_{i}$, stress $\sigma_i$, and leak-off $C'_{i}$. Let the element size be the same and equal to $h$. Two fracture propagation directions are considered, ``upward'' and ``downward'', as indicated in the figure. The distance from the front, $s$, is measured from the fracture front location (indicated by the red horizontal lines) towards inside of the fracture. Finally, the fill ratio, $f$, indicates the fraction of the filled element, and is related to the fracture front position and element size as indicated in the figure. 

The fracture front is determined by the fill ratio and the direction of propagation. For any given fill ratio, it is possible to compute the apparent fracture toughness as
\begin{equation}\label{apparentKC}
    K'_{u,d}(f)=K'(f)+\dfrac{8}{\pi}\int_0^{(f\!+\!1)h} \dfrac{\sigma(s)-\bar\sigma_{u,d}(s)}{\sqrt{s}}\,ds.
\end{equation}
Here $K'_{u,d}$ is the apparent scaled fracture toughness in either upward or downward direction of fracture growth. They both depend on the fill ratio $f$. Note that the definition of $s$ is different for the upward and downward fracture directions, which causes a difference between the integration for these two cases.

The apparent fracture toughness consists of the material fracture toughness that varies versus fill ratio due to layering and contribution of the stress layering. The latter comes from the expression for the stress intensity factor for a loaded semi-infinite fracture~\cite{Rice1968}. The loading is the difference between the actual layered stress and the average stress or the stress that is assigned to the given element. The integration is performed over the ``main'' element as well as the previous element, that can be an element above or below depending on the fracture propagation direction. This is done to increase accuracy for small values of the fill ratio. The average stress is computed according to
\begin{equation}\label{averSigma}
\bar \sigma_{u,d}(s) = \left\{\begin{aligned}
&\sigma^t_{u,d},\quad 0\leqslant s\leqslant hf,\\
&\sigma^p_{u,d},\quad hf\leqslant s\leqslant (f\!+\!1)h.
\end{aligned}\right.\quad \sigma^t_{u,d} = \sigma^p_{u,d}+\Delta\sigma_{u,d} =  \dfrac{1}{hf}\int_0^{hf}\!\!\!\!\!\sigma(s')\,ds',\quad \sigma^p_{u,d}=\dfrac{1}{h}\int_{hf}^{(f\!+\!1)h}\!\!\!\!\! \sigma(s') \, ds'.
\end{equation}
In the above expression, stress averaged over the open portion of the fracture is used for the tip element $\sigma^t_{u,d}$ (this value varies with fill ratio) and the average stress is used for the ``previous'' element $\sigma^p_{u,d}$, that can be either above or below depending on the fracture direction. Note that in this case the stress difference between the elements is $\Delta\sigma_{u,d} = \sigma^t_{u,d}\!-\!\sigma^p_{u,d}$ and it varies with the fill ratio.

To compute leak-off at the tip element, the leak-off rate is integrated over the element as 
\begin{equation}\label{tipleakoff}
    q_{tip}=\dfrac{1}{h}\int_0^{hf} \dfrac{C'(s)\,ds}{\sqrt{t\!-\!t_0(s)}} = \dfrac{1}{h}\int_0^{h(f-f_0)} \dfrac{C'(s)\,ds}{\sqrt{t\!-\!t_0(s)}}+\dfrac{1}{h}\int_{h(f-f_0)}^{hf} \dfrac{C'(s)\,ds}{\sqrt{t\!-\!t_0(s)}}.
\end{equation}
In the latter expression, the integral is split into two, one that accounts for the contribution of the existing fracture and another one due to newly created surface area. By assuming that the fracture tip propagates with a constant velocity $\dot s$, the contribution of the newly created fracture can be further simplified as
\begin{equation}\label{tipleakoff2}
\dfrac{1}{h}\int_{0}^{h(f-f_0)} \dfrac{C'(s)\,ds}{\sqrt{t\!-\!t_0(s)}} = \dfrac{\sqrt{\dot s}}{h}\int_{0}^{h(f-f_0)} \dfrac{C'(s)\,ds}{\sqrt{s}}\approx \sqrt{\dfrac{4\dot s(f\!-\!f_0)}{h}} \dfrac{1}{h(f\!-\!f_0)}\int_{0}^{h(f-f_0)} C'(s)\,ds.
\end{equation}
Here the last step assumes that the leak-off coefficient does not vary significantly over the newly generated fracture area and hence the average is used. By introducing the average leak-off coefficient in the upward and downward directions as
\begin{equation}\label{leakav}
 \bar C'_{u,d}(f) = \dfrac{1}{fh}\int_0^{fh} C'(s)\,ds,
\end{equation}
the leak-off rate from the tip element~(\ref{tipleakoff}) reduces to
\begin{equation}\label{tipleakofffinal}
    q_{tip} = \dfrac{1}{h}\int_{0}^{hf_0} \dfrac{C'(s')\,ds'}{\sqrt{t\!-\!t_0(s')}}+\sqrt{\dfrac{4\dot s(f\!-\!f_0)}{h}} \dfrac{f \bar C'_{u,d}(f)-f_0 \bar C'_{u,d}(f_0)}{f-f_0},
\end{equation}
where the new coordinate $s'=s-h(f\!-\!f_0)$ is introduced in lieu if $s$ in the integral to highlight the fact it does not depend on $f$. Note that the integral in the above expression is computed numerically once per time step, while the second part can be evaluated multiple times to ``probe'' the new fracture front location. 

Next, an effective leak-off coefficient is needed for the tip asymptotic solution $\tilde w$. In order to obtain one, note that the total leak-off flux for the case of a steadily propagating fracture in a homogeneous formation is $q_{tip}=C_{tip}\sqrt{4\dot s f/h}$, see e.g.~(\ref{tipleakoff2}). Thus, the effective leak-off coefficient for the tip element can be computed as
\begin{equation}\label{Ctip}
    C_{tip} = \sqrt{\dfrac{h}{4\dot s f}}\,q_{tip}.
\end{equation}
Recall that the tip velocity is computed as $\dot s=h(f-f_0)/\Delta t$, where $\Delta t$ is the time step. While this is an approximation, it preserves the total leak-off related fluid velocity at the element edge, which is expected to provide reasonably accurate results.


To summarize, in order to account for layers, the fracture front evolution equations~(\ref{bc_visc}) are modified as
\begin{equation}\label{bc_visc_lay}
      w_{tip} = \dfrac{2K'_{u,d}(f)\tilde w(f,f_0)}{3E'}f^{3/2} h^{1/2}+\dfrac{8\Delta\sigma_{u,d}(f) h f^2}{3\pi E'},\quad \Delta\sigma_f = \dfrac{K'_{u,d}(f)\tilde w(f,f_0)}{h^{1/2}}\Sigma_K(f)+\Delta\sigma_{u,d}(f) \Sigma_S(f),
\end{equation}
where the apparent toughness~(\ref{apparentKC}) and leak-off~(\ref{Ctip}) are also used in the function $\tilde w$. Equation~(\ref{bc_visc_lay}) is essentially the same as the one for homogeneous formations~(\ref{bc_visc}), but utilizes local values of stress, toughness, and leak-off as a function of the fill ratio or fracture front position. Note that the total stress assigned for the tip element is calculated as
\begin{equation}\label{tip_stress}
\sigma_{tip} = \sigma^p_{u,d}+\Delta\sigma_{f},
\end{equation}
i.e. is equal to the average stress in the previous element plus the fictitious stress. Finally, the leak-off flux from the tip element is computed using~(\ref{tipleakofffinal}). This approach relies heavily on analytical expressions and approximations to ensure that the resulting algorithm is computationally efficient. These approximations will then be effectively tested during benchmarking of the algorithm. Given the unique ability of the algorithm to capture the effect of Multiple Layers for Tip Elements, it is called ``MuLTipEl".

\begin{figure}
\centering \includegraphics[width=1\linewidth]{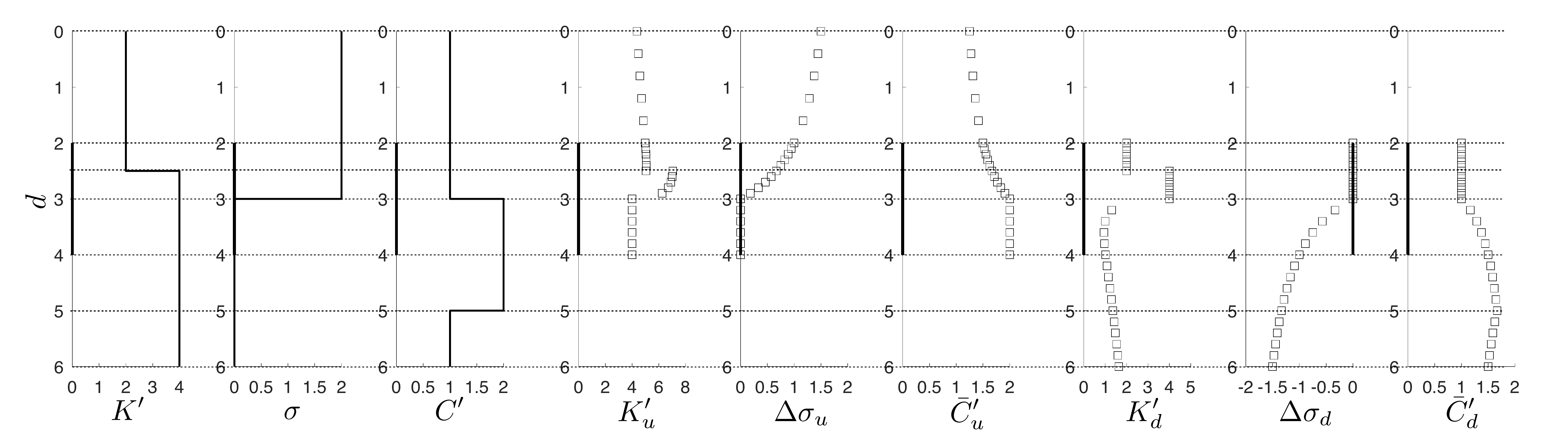}
\caption{An example variation of properties within the element and its neighbours. The first three tracks show variation of toughness, stress, and leak-off versus depth. The second three tracks show the apparent toughness, stress difference, and average leak-off for the upwards fracture propagation. The last three tracks show the apparent toughness, stress difference, and average leak-off for the downwards fracture propagation.}
\label{fig5}
\end{figure}
To illustrate behavior of the apparent properties used for the upward and downward fracture propagation, Fig.~\ref{fig5} considers an example variation of toughness, stress and leak-off versus depth for the ``main'' element and its neighbours above and below. In particular, the first track shows the scaled fracture toughness versus depth, the second track shows the variation of stress, and the third track shows scaled leak-off. Units are not relevant for this example and are therefore omitted. The ``main'' element is highlighted by the bold black line on every plot and the layer boundaries are shown by the dashed lines. The last six tracks show the variation of the apparent toughness, stress difference, and average leak-off for the upward and downward propagation directions that are computed using~(\ref{apparentKC}), (\ref{averSigma}), and~(\ref{leakav}). 

To facilitate numerical computations, each layer is subdivided into several points (six in this example). The first point is always located at the layer top and the last point is at the layer bottom. The rest of the points are uniformly distributed along the layer thickness. All the relevant quantities, such as $K'_{u,d}$, $\bar C'_{u,d}$, and $\Delta \sigma_{u,d}$, are pre-computed at these evaluation points and are stored for future use. Note that the actual value of the leak-off coefficient as well as an array of ``open'' times $t_0$ are also stored at these points to compute leak-off from the tip element using~(\ref{tipleakofffinal}). With such a discretization, each layer boundary has two points, one with the properties of the layer above and another one with the properties of the layer below. This is included to capture discontinuous behavior of fracture toughness. Also, there are several points per layer to capture non-linear behavior of the quantities under consideration. While linear interpolation can be used in between the points. The element boundaries are also considered to be layer boundaries for convenience. 

Since the upward propagating fracture can occupy either the main element or the one above it, the evaluation points of the fill ratio (and hence depth) span these two elements. Similarly, for downward propagation, the evaluation points are located inside the primary element and the one below. Also, the variation of quantities within the element is very different for the upward and downward propagation, which highlights that it is essential to differentiate between the two.

For the upward growth, starting from the depth of $d=4$ and above, all the quantities have initially constant values. Once the first layer boundary at $d=3$ is reached, the average stress difference between the ``main'' element and the one below starts to accumulate, while the average leak-off reduces, which is consistent with the layered rock properties. Also, to compensate for the stress variation, the effective toughness increases rapidly due to entering the stress barrier. Next layer is reached at $d=2.5$. This layer is characterized by the drop of fracture toughness. Therefore, stress and leak-off continue having monotonic behavior, while toughness exhibits a jump. After that, the properties are constant, there are no more layers, and therefore all quantities proceed with monotonic behavior. 

Situation with downward propagation is different. Again, the starting point is constant behavior from $d=2$ until $d=2.5$. Then toughness jumps, but remains constant. From $d=3$ and downwards, toughness and stress vary continuously since the fracture tip enters the stress drop region. Leak-off increases gradually until $d=5$ and then starts to decrease since the leak-off gets reduced after $d=5$.

\section{Numerical examples}\label{SecNum}

\begin{figure}
\centering \includegraphics[width=0.4\linewidth]{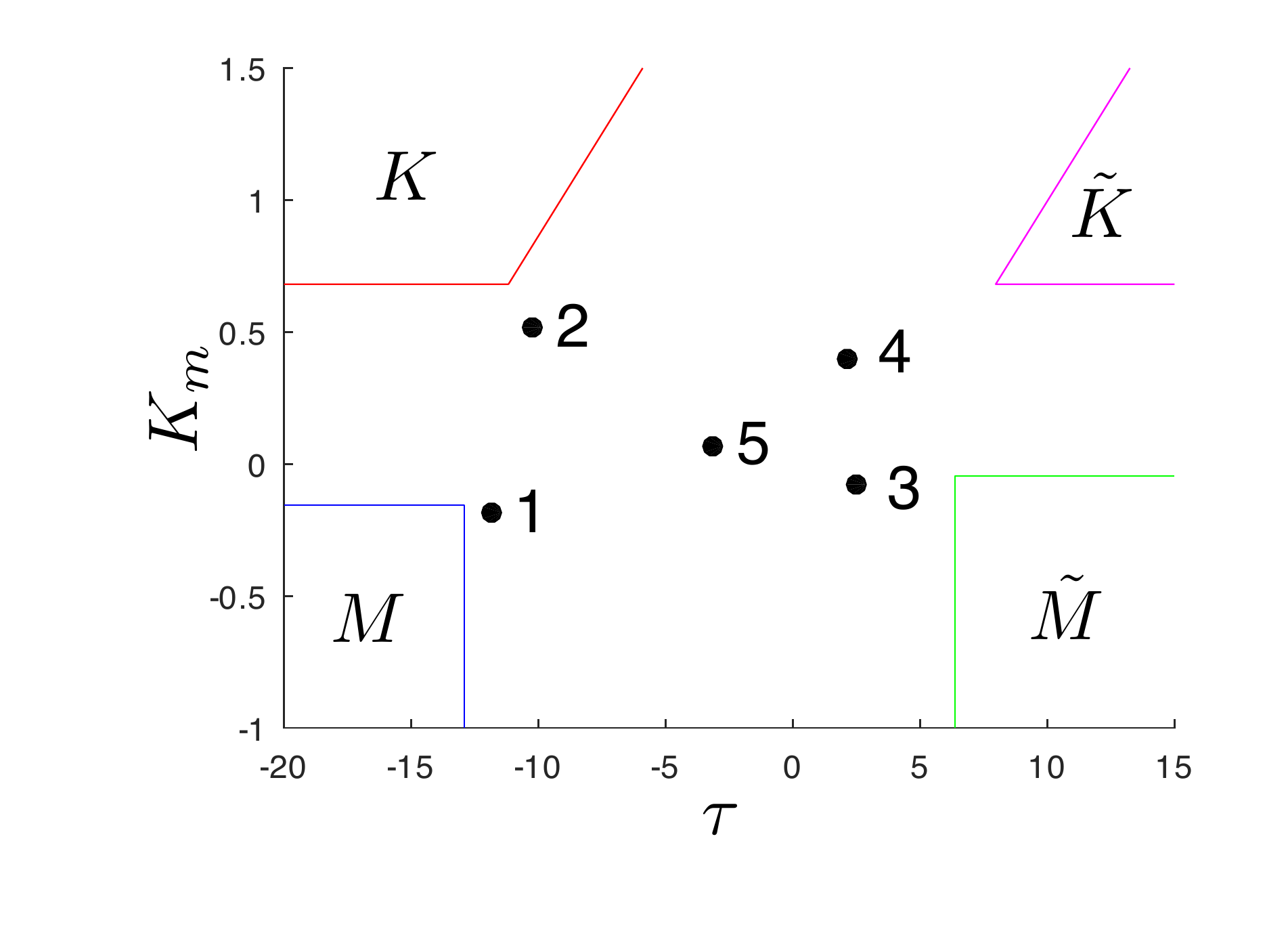}
\vspace*{-5mm}
\caption{Parametric space for the problem of a plane strain hydraulic fracture in homogeneous formation. Regions of applicability of the limiting solutions are outlined by the colored lines (see text for description). Location of the example problem parameters are shown by black circular markers, which are enumerated for convenience.}
\label{fig6}
\end{figure}
To illustrate ability of the approach to smoothly track the fracture front, propagation of a plane strain fracture in homogeneous formation is considered first. The parametric space for the problem is shown in Fig.~\ref{fig6}. There are two dimensionless parameters that determine the solution~\cite{Hu2010,Dont2017c}
\begin{equation}\label{dimpar}
    K_m = \Bigl(\dfrac{K'H}{\mu'E'^3Q_0}\Bigr)^{1/4},\qquad \tau = \dfrac{t E'C'^6H^3}{\mu'Q_0^3},
\end{equation}
where the first one is the dimensionless toughness and the second one is the dimensionless time. There are four limiting regimes shown in Fig.~\ref{fig6}: storage-viscosity or $M$ (blue lines), storage-toughness or $K$ (red lines), leak-off-viscosity or $\tilde M$ (green lines), and leak-off-toughness or $\tilde K$ (magenta lines). These regimes quantify competition between toughness and viscosity for the dissipation mechanism as well as fluid storage inside the fracture or in the formation. In order to probe the parametric space in different locations, the following 5 sets of parameters are considered:
\begin{enumerate}
     \item $K_{Ic}=4$~MPa$\cdot$m$^{1/2}$, $C_l=2\times10^{-6}$~m/s$^{1/2}$, $\mu=0.4$~Pa$\cdot$s, $Q_0=1$~m$^3$/s;
     \item $K_{Ic}=8$~MPa$\cdot$m$^{1/2}$, $C_l=2\times10^{-6}$~m/s$^{1/2}$, $\mu=0.01$~Pa$\cdot$s, $Q_0=1$~m$^3$/s;
     \item $K_{Ic}=4$~MPa$\cdot$m$^{1/2}$, $C_l=6\times10^{-4}$~m/s$^{1/2}$, $\mu=0.05$~Pa$\cdot$s, $Q_0=3$~m$^3$/s;
     \item $K_{Ic}=8$~MPa$\cdot$m$^{1/2}$, $C_l=4\times10^{-4}$~m/s$^{1/2}$, $\mu=0.01$~Pa$\cdot$s, $Q_0=3$~m$^3$/s;
     \item $K_{Ic}=6$~MPa$\cdot$m$^{1/2}$, $C_l=5\times10^{-5}$~m/s$^{1/2}$, $\mu=0.2$~Pa$\cdot$s, $Q_0=1$~m$^3$/s.
\end{enumerate}
The common parameters are: $H=300$~m, $E=20$~GPa, $\nu=0.2$, and pump time $t_{pump} = 50$~min. Element size of $h=50$~m and the time step $\Delta t=10$~s are used for all cases. Location of these parameter sets in the parametric space is shown in Fig.~\ref{fig6} by black circular markers. As can be seen from the figure, the first four sets of parameters are located close to the limiting solutions, while the fifths point represents an intermediate case.

\begin{figure}
\centering \includegraphics[width=0.9\linewidth]{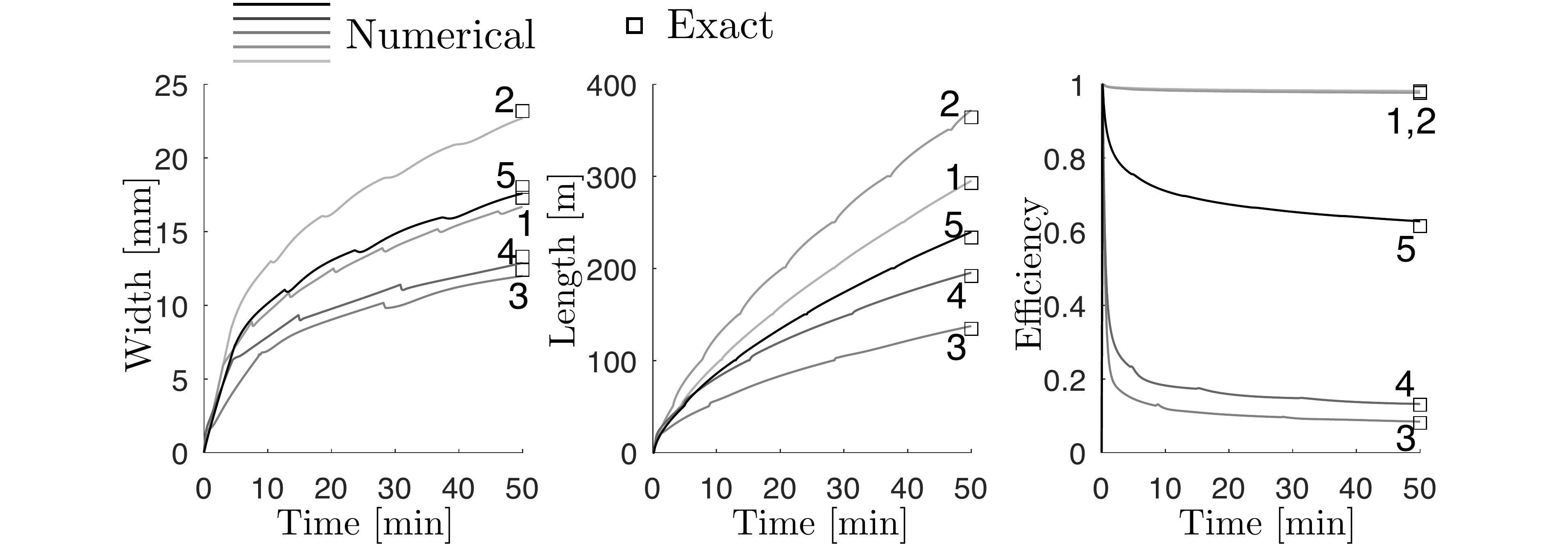}
\caption{Benchmarking of the numerical solution against the reference solution for the case of a plane strain hydraulic fracture in a homogeneous formation. Five cases are considered, with the location in the parametric space shown in Fig.~\ref{fig6}. Solid lines show numerical solution and square markers show the reference or exact solution. The left panel shows width at the wellbore versus time. The middle panel compares length of one fracture wing. The right panel plots efficiency versus time.}
\label{fig7}
\end{figure}
Results of numerical simulations are depicted in Fig.~\ref{fig7}, see Appendix~\ref{app2} for the numerical scheme. The left panel plots fracture width at the wellbore versus time, the middle panel shows fracture length (for one fracture wing) versus time, while the right panel shows time history of efficiency defined as the ratio between the fracture volume and the injected volume. Solid lines show results of the numerical simulations, while square markers indicate the reference solution~\cite{Dont2017c} for the corresponding parameters. All the lines are enumerated according to the parameters specified above and the location in the parametric space shown in Fig.~\ref{fig6}. As can be seen from the results, the fictitious stress algorithm allows to smoothly track the fracture front and to match the reference solution for all cases. The fracture length has some visible features occurring during initialization of a new element, but they are relatively small. The fracture width has some more noticeable changes. They are consistent with earlier observations stating that the width at the wellbore can drop after initialization of a new element, see Fig.~\ref{fig3} and the corresponding description in the text. The fracture efficiency or volume is also matching for all cases. Note that a relatively coarse mesh is used in this example (three to eight elements per fracture half length) to highlight accuracy of the solution on a coarse mesh and to illustrate solution features during opening of a new element.

To evaluate accuracy of the developed numerical algorithm under various layered formations, several numerical examples are considered next. Solution is computed three times for each example using different element size: 100~m, 50~m, and 25~m. Note that the time step is 40~s, 20~s, and 10~s for these three meshes, respectively. The following parameters are common for all cases (unless specifically stated otherwise): $E=20$~GPa, $\nu=0.2$, $Q_0=0.5$~m$^3$/s, $H=300$~m, $t_{pump} = 50$~min, and $\mu=0.02$~Pa$\cdot$s.

\begin{figure}
\centering \includegraphics[width=1.1\linewidth]{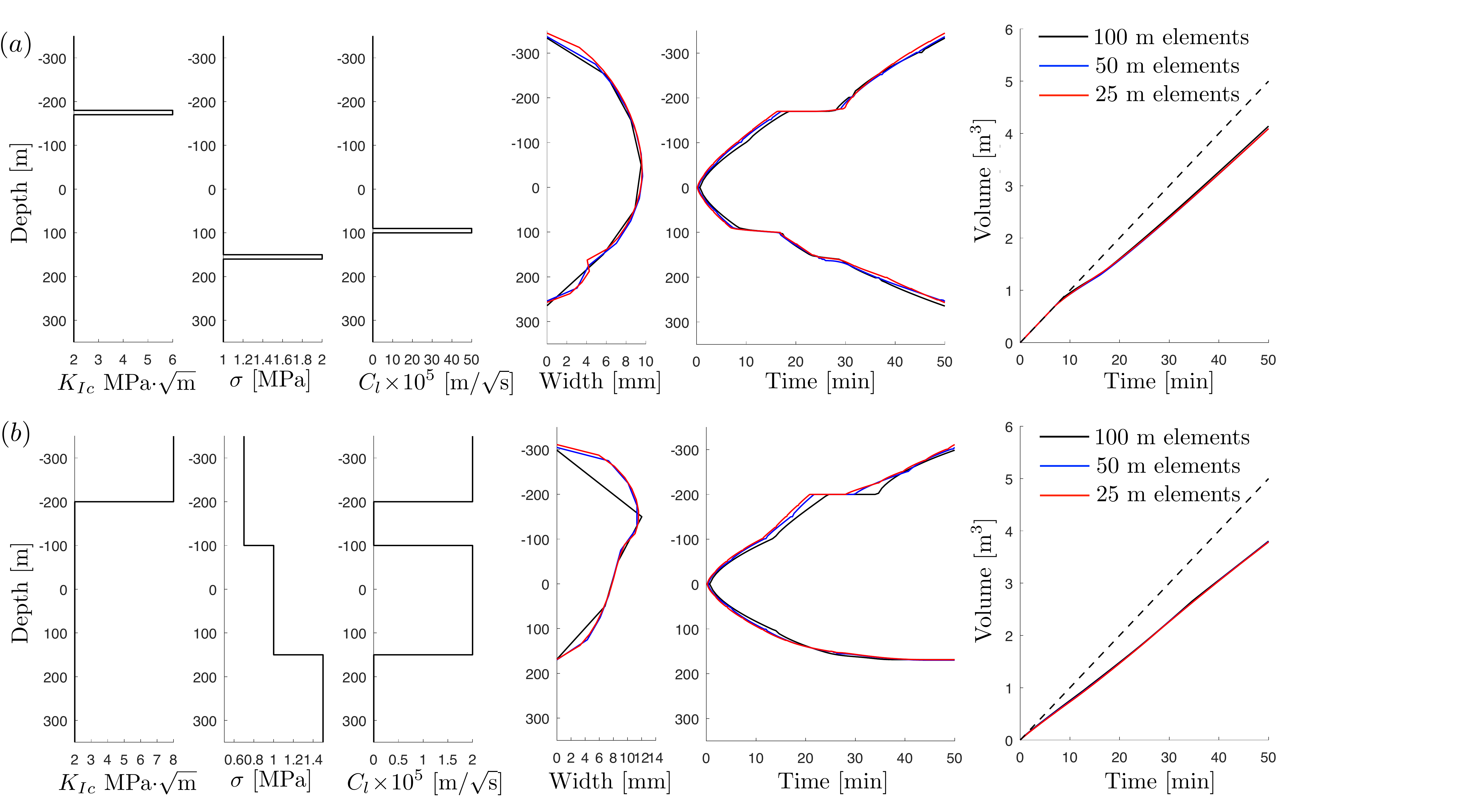} \hspace*{-2cm}
\caption{Results of numerical simulations of hydraulic fracture growth in a formation with thin $(a)$ and thick $(b)$ layers. The left three tracks show variation of toughness, stress, and leak-off coefficient versus depth. The fourth track shows the fracture width at the final time. The fifth track shows evolution of fracture length versus time, and the last track shows the fracture volume versus time. The solid black, blue, and red lines show results computed using 100~m, 50~m, and 25~m element size, respectively. The dashed black line shows the injected volume.}
\label{fig8}
\end{figure}

Fig.~\ref{fig8}$(a)$ shows an example of a hydraulic fracture growing through thin layers. In particular, three thin barriers of toughness, stress, and leak-off are considered, see the first three tracks in Fig.~\ref{fig8}$(a)$. Thickness of each of them is 10~m, which is smaller than the element size used. Results for the final fracture width, time evolution of fracture length and volume are shown by different colors for different meshes. There is a very good agreement between all the quantities. Also, one can clearly observe how fracture growth gets significantly suppressed by each of the barriers for a certain period of time, and then the fracture continues to grow normally after crossing the barrier. Regardless of the element size, the fracture is arrested exactly at the location of the thin layers. In addition, calculation of fluid leak-off is also accurate since the time history of fracture volume matches well. The dashed black line shows the injected volume. At approximately 10 minutes, the fracture reaches the leak-off layer and the fracture volume starts to deviate from the injected volume. The leak-off layer is thin, but has relatively large magnitude, so that a noticeable amount of fluid leaks into the formation. Finally, it is worth noting small differences of the fracture width profile, especially near the stress barrier. Solution computed on a coarse mesh is unable to capture fine details of the width variation, but instead provides an accurate average value.

Fig.~\ref{fig8}$(b)$ shows results computed for the case of thick layers, in which layer thickness is comparable to the element size. As for the previous case, the first three tracks on the figures show the input rock properties, while the final width, time evolution of length and volume are shown further to the right. The fracture first grows symmetrically, and then the upper tip reaches a stress drop and a leak-off drop, which promotes upwards growth. After that, the fracture reaches a toughness and leak-off barrier at the top and a stress barrier at the bottom. It eventually crosses the toughness barrier and propagates further up. As for the previous case, results computed using three different meshes are very consistent. The fracture width seems to be less accurate near the fracture tips, but this is also caused by visualization, in which element centers are connected by straight lines rather than being drawn as piece-wise constant. To further confirm this, the fracture volumes are nearly identical for all the cases, even though based on the width profile, the fracture obtained with 100~m elements seems to have a smaller volume. 

\begin{figure}
\centering \includegraphics[width=1.1\linewidth]{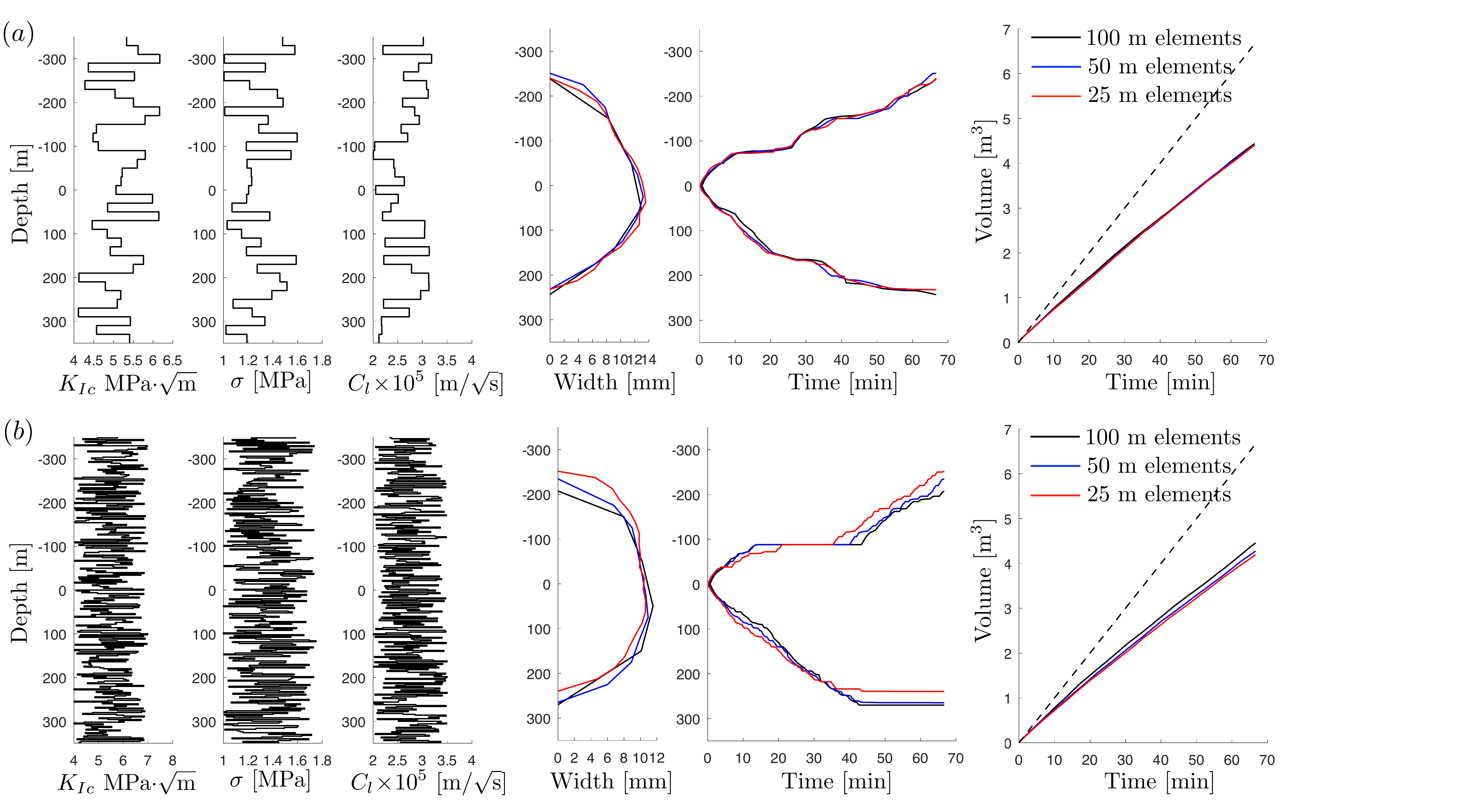} \hspace*{-2cm}
\caption{Results of numerical simulations of hydraulic fracture growth in a formation with randomly generated layers with size 20~m $(a)$ and 2~m $(b)$. The left three tracks show variation of toughness, stress, and leak-off coefficient versus depth. The fourth track shows the fracture width at the final time. The fifth track shows evolution of fracture length versus time, and the last track shows the fracture volume versus time. The solid black, blue, and red lines show results computed using 100~m, 50~m, and 25~m element size, respectively. The dashed black line shows the injected volume.}
\label{fig9}
\end{figure}

To demonstrate ability of the algorithm to tackle large amount layers, Fig.~\ref{fig9} shows the results for randomly generated rock properties. In particular, Fig.~\ref{fig9}$(a)$ shows the results with 20~m layers and Fig.~\ref{fig9}$(b)$ shows the results with 2~m layers. Keep in mind that the element sizes considered are the same as before, namely, 100~m, 50~m, and 25~m, which makes the layer size noticeably smaller than the element size. Also note that the pump time for these problems is increased to $t_{pump} = 66.7$~min. Similarly to the previous cases, the results show a very good degree of consistency for all meshes, albeit the differences are larger for smaller layers. These examples represent relatively good cases, while it is possible to find situations, in which the agreement is not as good. For instance, there are situations, in which the barriers on the top and at the bottom are very similar. However, once the fracture grows through one of the barriers, it does not have enough energy to cross the second one. Depending on the mesh used, the solution can either trigger growing through one or another barrier, which leads to very different results. The more layers are in the system, the higher chance is to encounter such a situation. Therefore, the case with 2~m layers is less accurate compared to the case with 20~m layers. Regarding computational time, no significant difference is observed between these cases since the solution for the tip element is computed analytically. Therefore, the developed approach allows to account for large amount of layers without significantly affecting the overall computational time of the algorithm.

\section{Summary}\label{SecSum}

This study presents the Multi Layer Tip Element (MuTipEl) algorithm for solving the problem of a plane strain hydraulic fracture propagating in a layered formation. In particular, toughness, stress, and leak-off layers are considered and the layer size can be smaller than the element size. The goal is to develop an algorithm that is capable of solving such a problem on a coarse fixed mesh, such that there is no significant mesh dependence.

In order to achieve the goal mentioned above, the fracture front tracking algorithm is developed. To prevent the partially filled tip elements (that contain the fracture front) to be fully open, an additional fictitious stress is applied. The magnitude of this stress is such that the volume of the tip element is equal to that of the tip asymptotic solution emanating from the true front location. This concept has been inspired by using analytical solutions for the semi-infinite toughness dominated fracture and then further calibrated against a numerically computed toughness dominated solution for a finite fracture. The main conclusion stemming from here is that it is possible to ``penalize'' the fracture tip element by an additional stress to mimic true fracture front location within this element.

The algorithm has been extended to capture the effects of viscosity and leak-off on the near-tip behavior. In order to do this, the concept of an apparent toughness is used, whereby the actual fracture toughness is replaced by the ratio between the universal asymptotic solution (that accounts for the viscosity and leak-off) and the corresponding toughness solution. Testing results for different regimes demonstrate that such a modification leads to accurate results, thus showing that the developed concept of front tracking approach works well for all problem parameters (and homogeneous rock properties). This demonstrates once again the need of using the near-tip solution to obtain accurate results on a coarse mesh.

The effect of thin layers is incorporated by using the effective values for stress, toughness, and leak-off for the tip element, which vary as a function of fracture front location. In particular, the average stress is used to preserve the force, local toughness is complemented by the stress intensity factor correction due to the difference between the actual and averaged stress, and leak-off is computed based on the history of fracture front evolution through the tip element and its values for different layers. Several numerical examples are presented, including the ones with thin layers, thick layers, and randomly generated layers. It is concluded that the developed algorithm is able to tackle layers accurately even if situations, in which the layer size is much smaller than the element size. At the same time, since analytical expressions are used to solve the problem of the tip element, there is no drastic influence of the number of layers on the overall computational performance of the algorithm.

To conclude, this study shows two main points: i) it is possible to track the fracture front by adding an additional fictitious stress to the tip elements to prevent them from being immediately wide open, and ii) in order to capture the effects of layers within the tip element, one essentially needs to solve the local problem of a semi-infinite hydraulic fracture propagating in a layered formation with a given tip volume. An approximate analytical solution is used for the latter problem for the purpose of computational efficiency, but it can conceptually be replaced by a numerical solution as well.

\section*{Acknowledgements}
I would like to thank Dr. Mark McClure (ResFrac Corporation) for useful discussions and for giving me the opportunity to develop this algorithm.

\appendix
\section{Calculation of the fictitious stress for a uniformly pressurized fracture}\label{app1}

The purpose of this Appendix is to derive the solution for~(\ref{wNum}) and~(\ref{voltipmatch}). Due to linearity of the problem, the contributions of stress barrier and toughness can be considered independently.

By assuming that $\Delta\sigma=0$, equations~(\ref{voltipmatch}) can be rewritten as
\begin{eqnarray}\label{voltipmatchK}
  \sum \bar{\boldsymbol{C}}^{-1} \boldsymbol{v_1}  \Pi_K + \sum \bar{\boldsymbol{C}}^{-1}  \boldsymbol{v_2} \Sigma_K  &=& \pi\Bigl(\dfrac{N\!+\!f}{2}\Bigr)^{3/2},\notag\\
  \boldsymbol{v_2}^T \bar{\boldsymbol{C}}^{-1} \boldsymbol{v_1}  \Pi_K + \boldsymbol{v_2}^T \bar{\boldsymbol{C}}^{-1}  \boldsymbol{v_2} \Sigma_K  &=& \dfrac{2}{3}f^{3/2}.
\end{eqnarray}
where the following dimensionless quantities are introduced
\begin{equation}\label{scalingK}
    \Sigma_K = \dfrac{h^{1/2}}{K'} \Delta\sigma_f,\qquad \Pi_K= \dfrac{h^{1/2}}{K'} p_0,\qquad \bar {\boldsymbol{C}} = \dfrac{h}{E'} \boldsymbol{C},\qquad \bar L = \dfrac{L}{h}.
\end{equation}

At the same time, the problem with $K'=0$ can be written in the dimensionless form as
\begin{eqnarray}\label{voltipmatchS}
  \sum \bar{\boldsymbol{C}}^{-1} \boldsymbol{v_1}  \Pi_S + \sum \bar{\boldsymbol{C}}^{-1}  \boldsymbol{v_2} \Sigma_S  &=& 4N\sqrt{(N\!+\!f)^2-N^2},\notag\\
  \boldsymbol{v_2}^T \bar{\boldsymbol{C}}^{-1} \boldsymbol{v_1}  \Pi_S + \boldsymbol{v_2}^T \bar{\boldsymbol{C}}^{-1}  \boldsymbol{v_2} \Sigma_S  &=& \dfrac{8f^2}{3\pi},
\end{eqnarray}
where the scaling is given by
\begin{equation}\label{scalingS}
    \Sigma_S = \dfrac{\Delta\sigma_f}{\Delta\sigma},\qquad \Pi_K= \dfrac{p_0}{\Delta\sigma},\qquad \bar {\boldsymbol{C}} = \dfrac{h}{E'} \boldsymbol{C},\qquad \bar L = \dfrac{L}{h}.
\end{equation}

The above systems of equations~(\ref{voltipmatchK}) and~(\ref{voltipmatchS}) can be solved for the dimensionless fracture pressures $\Pi_K$ and $\Pi_S$, as well as for the dimensionless fictitious stresses $\Sigma_K$ and $\Sigma_S$ for the given fill ratio $f$ and number of elements $N$. Therefore, the total solution for the fictitious stress is 
\begin{equation*}
    \Delta\sigma_f = \dfrac{K'}{h^{1/2}}\Sigma_K(f,N)+\Delta\sigma \Sigma_S(f,N),
\end{equation*}
which is obtained by using the scaling factors from~(\ref{scalingK}) and~(\ref{scalingS}).

\section{Numerical algorithm}\label{app2}

Numerical algorithm for the plane strain hydraulic fracture problem under consideration is based on the fixed mesh, characterized by the element size $h$ and total number of elements $N_E$. Element centers are located at $z_j=h(j\!-\!j_{inj}\!-\!1/2)$, $j=1..N_E$, where the index $j_{inj}$ determines location of the injection points, namely the injection is divided evenly between two elements with indices $[j_{inj},j_{inj}+1]$. The fracture initializes with two elements located at $-h/2$ and $h/2$, while the injection point $j_{inj}=1$. The arrays of widths and pressures, defined at points $z_j$ and evaluated at time $t_i$, are denoted by ${\boldsymbol w}^i$ and ${\boldsymbol p}^i$, respectively.

With the above definitions, equations~(\ref{volumebalance}) and~(\ref{elas}) can be discretized using backward time stepping as
\begin{equation}\label{eqdiscr}
\dfrac{{\boldsymbol w}^{i+1}-{\boldsymbol w}^{i}}{\Delta t}={\boldsymbol A}({\boldsymbol w}^{i+1}) \boldsymbol{p}^{i+1}+{\boldsymbol S}^i,\qquad {\boldsymbol p}^{i+1} = {\boldsymbol \sigma}^i+{\boldsymbol C}{\boldsymbol w}^{i+1},\qquad \Delta t = t_{i+1}-t_i,
\end{equation}
where the matrix ${\boldsymbol A}({\boldsymbol w}^{i+1})$ approximates the flux derivative using central difference as
\begin{equation}\label{Adef}
[{\boldsymbol A}({\boldsymbol w}^{i+1}) \boldsymbol{p}^{i+1}]_j =\dfrac{1}{\mu' h^2} \Bigl( \lambda^{i+1}_{j+1/2} (p^{i+1}_{j+1}\!-\!p^{i+1}_{j})-\lambda^{i+1}_{j-1/2} (p^{i+1}_{j}\!-\!p^{i+1}_{j-1})\Bigr),\qquad \lambda^{i+1}_{j\pm1/2} = \dfrac{(w^{i+1})^3_j+(w^i)^3_{j\pm1}}{2},
\end{equation}
the elasticity matrix is given by
\begin{equation}\label{Cdiscr}
{\boldsymbol C}_{ij} = \dfrac{E'}{4\pi}\Bigl( \dfrac{1}{z_j\!-\!z_i\!+\!h/2}-\dfrac{1}{z_j\!-\!z_i\!-\!h/2}\Bigr),
\end{equation}
and the source term is discretized as
\begin{eqnarray}\label{Sdiscr}
{\boldsymbol S}^i_j &=& -\dfrac{2 C'_j }{\Delta t}\Bigl(\sqrt{t_{i+1}\!-\!t_{0,j}}-\sqrt{t_i\!-\!t_{0,j}}\Bigr)+\dfrac{Q_0}{2hH}\delta_{j,j_{inj}}+\dfrac{Q_0}{2hH}\delta_{j,j_{inj}+1},\qquad j=2...N_E-1,\notag\\
{\boldsymbol S}^i_1 &=& -q^i_{u,tip},\qquad {\boldsymbol S}^i_{N_E} = -q^i_{d,tip}.
\end{eqnarray}
Here $C'_j$ represents leak-off coefficient averaged over the $j$th cell, $t_{0,j}$ is the open time for the $j$th element computed as an average between the element initiation time and the time at which the element becomes fully open, ``$\delta_{ij}$'' denotes Kronecker delta, while $q_{u,tip}$ and $q_{d,tip}$ represent leak-off from tip elements in the upward and downward directions. Finally, ${\boldsymbol \sigma}^i$ is the vector of stresses defined as
\begin{equation}\label{sigmadiscr}
{\boldsymbol \sigma}^i_j = \sigma_j,\qquad {\boldsymbol \sigma}^i_1 = \sigma_{u,tip},\qquad {\boldsymbol \sigma}^i_{N_E} = \sigma_{d,tip},
\end{equation}
where $\sigma_j$ is the average stress acting on $j$th element, while $\sigma_{u,tip}$ and $\sigma_{d,tip}$ represent tip stresses for the upward and downward propagation. Once the tip leak-off and stress values are known, the system of nonlinear equations~(\ref{eqdiscr})--(\ref{sigmadiscr}) is solved iteratively for every time step. 

As can be seen from the above equations, the coupling between the tip elements and the rest of the fracture comes solely from the definitions of the tip leak-offs and stresses. The latter depend on the value of the fill ratio in each element. It is therefore instructive to briefly describe numerical algorithm associated with the fracture front location. First, at the initialization of a new element, the quantities $K'_{u,d}(f)$, $\Delta \sigma_{u,d}(f)$, and $C'_{u,d}(f)$ are pre-computed at the evaluation points for the given layered properties, see e.g. Fig.~\ref{fig5}. Also, the value of leak-off coefficient is stored and the array of initialization times $t_0$ are populated at the evaluation points to facilitate computation of leak-off from the tip element~(\ref{tipleakofffinal}). In addition to that, average values for stress and leak-off coefficient are stored for future use in the ``main'' fracture. The first equation in~(\ref{bc_visc_lay}) determines new fill ratio for the given tip width, and it can be written as
\begin{eqnarray}\label{wtipud}
      {\boldsymbol w}^{i}_1 &=& \dfrac{2K'_{u}(f^{i+1}_u)\,\tilde w(f^{i+1}_u,f^{i}_{u})}{3E'}(f_u^{i+1})^{3/2} h^{1/2}+\dfrac{8\Delta\sigma_{u}(f^{i+1}_u)\, h\, (f_u^{i+1})^2}{3\pi E'},\notag\\
      {\boldsymbol w}^{i}_{N_E} &=& \dfrac{2K'_{d}(f^{i+1}_d)\,\tilde w(f^{i+1}_d,f^{i}_{d})}{3E'}(f_d^{i+1})^{3/2} h^{1/2}+\dfrac{8\Delta\sigma_{d}(f^{i+1}_d)\, h\, (f_d^{i+1})^2}{3\pi E'}.
\end{eqnarray}
To solve the above equations for the new fill ratio, they are first evaluated at the layer boundaries to find a layer, in which the fracture is arrested. This procedure starts from the layer in which the fracture front was located at the previous time step. Then, the same equations are evaluated to determine the sub-layer within which the fracture is arrested (recall that each layer is subdivided into several sublayers in order to capture the non-linear behavior of effective rock properties, see Fig.~\ref{fig5}). After this, the properties associated with layering are linearly interpolated between the neighbouring points and the new fill ratio is computed using bi-section method.

Once the fill ratio is computed, the tip stresses are evaluated using the second equation in~(\ref{bc_visc_lay}) and~(\ref{tip_stress}) to obtain
\begin{eqnarray}\label{sigmatipud}
    \sigma^i_{u,tip} &=& \sigma_2+\dfrac{K'_{u}(f^{i+1}_u)\tilde w(f^{i+1}_u,f^{i}_{u})}{h^{1/2}}\Sigma_K(f^{i+1}_u)+\Delta\sigma_{u}(f^{i+1}_u) \Sigma_S(f^{i+1}_u),\notag\\
    \sigma^i_{d,tip} &=& \sigma_{N_E-1}+\dfrac{K'_{d}(f^{i+1}_d)\tilde w(f^{i+1}_d,f^{i}_{d})}{h^{1/2}}\Sigma_K(f^{i+1}_d)+\Delta\sigma_{d}(f^{i+1}_d) \Sigma_S(f^{i+1}_d).
\end{eqnarray}
Finally, tip leak-off is evaluated using~(\ref{tipleakofffinal}) as
\begin{eqnarray}\label{tipleakoffud}
    q_{u,tip} &=& \sum \dfrac{ 2C'_{u,e} \Delta s_e}{h \Delta t} \Bigl(\sqrt{t_{i+1}\!-\!t_{0,u,e}}-\sqrt{t_i\!-\!t_{0,u,e}}\Bigr)+\dfrac{2 }{\sqrt{\Delta t}} \bigl(f^{i+1}_u \bar C'_{u}(f^{i+1}_u)-f_u^{i} \bar C'_{u}(f_u^{i})\bigr),\\
    q_{d,tip} &=& \sum \dfrac{ 2C'_{d,e} \Delta s_e}{h \Delta t} \Bigl(\sqrt{t_{i+1}\!-\!t_{0,d,e}}-\sqrt{t_i\!-\!t_{0,d,e}}\Bigr)+\dfrac{2 }{\sqrt{\Delta t}} \bigl(f^{i+1}_d \bar C'_{d}(f^{i+1}_d)-f_d^{i} \bar C'_{d}(f_d^{i})\bigr),
\end{eqnarray}
where $C'_{u,e}$ and $C'_{d,e}$ are the values of the leak-off coefficient at the evaluation points (recall that there are several evaluation points per each layer), $t_{0,u,e}$ and $t_{0,d,e}$ are the corresponding open times at these points, and the summation is computed from the element entrance edge to the fill ratio at the time instance $i$.

Once the fill ratio exceeds one in either upward or downward directions, a new element is initiated. The width is initialized as
\begin{equation}\label{widthupdate}
    {\boldsymbol w}^{i}_0 = \dfrac{(f_u^{i+1}-1)^{3/2}}{(f_u^{i+1})^{3/2}}{\boldsymbol w}^{i}_1,\qquad {\boldsymbol w}^{i}_{N_E+1} =\dfrac{(f_d^{i+1}-1)^{3/2}}{(f_d^{i+1})^{3/2}}{\boldsymbol w}^{i}_{N_E},
\end{equation}
The above expression implicitly assumes square root behavior near the tip, which is an assumption. However, if one uses the ``true" asymptote for this calculation, the difference will be relatively small. The tip stresses become
\begin{eqnarray}\label{sigmatipudUpdate}
    \sigma^i_{u,tip} &=& \sigma_1+\dfrac{K'_{u}(f^{i+1}_u)\tilde w(f^{i+1}_u,f^{i}_{u})}{h^{1/2}}\Sigma_K(f^{i+1}_u\!-\!1)+\Delta\sigma_{u}' \Sigma_S(f^{i+1}_u\!-\!1),\notag\\
    \sigma^i_{d,tip} &=& \sigma_{N_E}+\dfrac{K'_{d}(f^{i+1}_d)\tilde w(f^{i+1}_d,f^{i}_{d})}{h^{1/2}}\Sigma_K(f^{i+1}_d\!-\!1)+\Delta\sigma_{d}' \Sigma_S(f^{i+1}_d\!-\!1).
\end{eqnarray}
where the new stress differences are computed based on the total force consideration as
\begin{eqnarray}\label{sigmatipudUpdate2}
    \Delta\sigma_{u}' &=&\dfrac{f^{i+1}_u}{f^{i+1}_u-1}\bigl(\sigma_2-\sigma_1+\Delta\sigma_u(f^{i+1}_u)\bigr),\notag\\ 
    \Delta\sigma_{d}' &=&\dfrac{f^{i+1}_d}{f^{i+1}_d-1}\bigl(\sigma_{N_E-1}-\sigma_{N_E}+\Delta\sigma_d(f^{i+1}_d)\bigr). 
\end{eqnarray}
The tip leak-off rates are updated assuming that the leak-off coefficient does not change significantly over the fracture growth within the time step as
\begin{eqnarray}\label{tipleakoffudUpdate}
    q_{u,tip} &=& \dfrac{2 }{\sqrt{\Delta t}} \sqrt{\dfrac{f_u^{i+1}-1}{f_u^{i+1}-f_u^i}}\bigl(f^{i+1}_u \bar C'_{u}(f^{i+1}_u)-f_u^{i} \bar C'_{u}(f_u^{i})\bigr),\notag\\
    q_{d,tip} &=& \dfrac{2 }{\sqrt{\Delta t}} \sqrt{\dfrac{f_d^{i+1}-1}{f_d^{i+1}-f_d^i}}\bigl(f^{i+1}_d \bar C'_{d}(f^{i+1}_d)-f_d^{i} \bar C'_{d}(f_d^{i})\bigr).
\end{eqnarray}
If the upward propagation occurs, the element numeration shifts such the upward tip element is always the first element in the arrays. Finally, when computing the flux between the tip element and its neighbour using~(\ref{Adef}), the width of the tip element should be scaled with $1/f$ since $w$ represents the average width, while the actual width needed for the flux is $w/f$. Note that if one assumes the square root behavior near the tip, then the actual width becomes $3w/(2f)$. Numerical results, however, show that the difference between the two approaches is relatively minor.


\end{document}